\documentclass[12pt]{article}

\pdfoutput=1

\usepackage{graphics}
\usepackage{epsfig}
\usepackage{amsfonts}
\usepackage{amssymb}
\usepackage{amsmath}
\usepackage{dsfont}
\usepackage{pifont}
\usepackage{bbm}
\usepackage{multirow}
\usepackage{latexsym}
\usepackage{verbatim}
\usepackage{mcite}
\usepackage{cite}
\usepackage{units}
\usepackage[all]{xy}
\usepackage{cancel}
\usepackage{slashed}

\newcommand{\be}{\begin{equation}}
\newcommand{\ee}{\end{equation}}
\newcommand{\ben}{\begin{displaymath}}
\newcommand{\een}{\end{displaymath}}
\newcommand{\bea}{\begin{eqnarray}}
\newcommand{\eea}{\end{eqnarray}}

\newcommand{\bean}{\begin{eqnarray*}}
\newcommand{\eean}{\end{eqnarray*}}

\DeclareMathAlphabet{\mathpzc}{OT1}{pzc}{m}{it}

\begin{document}
\pagestyle{plain}


\makeatletter \@addtoreset{equation}{section} \makeatother
\renewcommand{\thesection}{\arabic{section}}
\renewcommand{\theequation}{\thesection.\arabic{equation}}
\renewcommand{\thefootnote}{\arabic{footnote}}


\setcounter{page}{1} \setcounter{footnote}{0}


\begin{titlepage}

\begin{flushright}
UUITP-01/15\\
\end{flushright}

\bigskip

\begin{center}

\vskip 0cm

{\LARGE \bf Geometric non-geometry} \\[6mm]

\vskip 0.5cm

{\bf Ulf Danielsson and\, Giuseppe Dibitetto \,  }\let\thefootnote\relax\footnote{{\tt ulf.danielsson@physics.uu.se, giuseppe.dibitetto@physics.uu.se}}\\

\vskip 25pt

{Institutionen f\"or fysik och astronomi, University of Uppsala, \\ Box 803, SE-751 08 Uppsala, Sweden \\[2mm]}

\vskip 0.8cm

\end{center}

\vskip 1cm

\begin{center}

{\bf ABSTRACT}\\[3ex]

\begin{minipage}{13cm}
\small

We consider a class of (orbifolds of) M-theory compactifications on $S^{d} \times T^{7-d}$ with gauge fluxes yielding minimally supersymmetric STU-models in 4D. We present a group-theoretical derivation of the corresponding flux-induced superpotentials and argue that the aforementioned backgrounds provide a (globally) geometric origin for 4D theories that only look locally geometric from the perspective of twisted tori. In particular, we show that Q-flux can be used to generate compactifications on $S^{4} \times T^{3}$.
We thus conclude that the effect of turning on non-geometric fluxes, at least when the section condition is solved, may be recovered by considering reductions on different topologies other than toroidal. 

\end{minipage}

\end{center}

\vfill

\end{titlepage}


\tableofcontents

\section{Introduction}
\label{sec:introduction}

The issue of studying compactifications of string theory producing satisfactory phenomenology has always been of utmost importance from several different perspectives. In particular, dimensional reductions of type IIA string theory and the possibility of generating a perturbative moduli potential induced by gauge fluxes and geometry has been widely explored in the literature over the last decade.

More specifically, type II reductions on twisted tori with gauge fluxes have received a lot of attention over the years owing to the possibility of analysing them in terms of their underlying lower-dimensional supergravity descriptions. In this context, a central role is played by those string backgrounds that can be described by a class of minimal supergravity theories \emph{a.k.a.} STU-models in four dimensions due to their remarkable simplicity. 

However, the search for (meta)stable de Sitter (dS) extrema within the above class of STU-models has turned out to be unsuccessful \cite{Hertzberg:2007wc,Caviezel:2008tf,Danielsson:2011au,Dibitetto:2011gm}. A possible further development of this research line includes the possibility of taking some strongly-coupled effects into account. Therefore, a very natural framework is that of M-theory compactifications. The corresponding flux-induced superpotentials present a complete set of quadratic couplings induced by the curvature \cite{Dall'Agata:2005fm}. Still, in such a context, reductions on twisted tori are known not to allow for any dS solutions either \cite{Derendinger:2014wwa}.

Within those STU-models describing M-theory on twisted tori, all the couplings higher than quadratic are still judged as non-geometric\cite{Shelton:2005cf}, \emph{i.e.} they do not admit any eleven-dimensional origin. Nevertheless, by moving to topologies other than toroidal, it is actually possible to find examples of flux superpotentials with homogenous degree higher than two. A particularly enlightening case is that of reduction on  $S^{7}$ yielding maximal $\textrm{SO}(8)$ gauged supergravity in four dimensions admitting a truncation to an STU-model featuring quartic superpotential couplings. Analytic continuations thereof describe non-compact gaugings exhibiting unstable dS extrema where, however, the internal space is non-compact \cite{Baron:2014bya}.

The goal of our work is to investigate which STU-models containing non-geometric fluxes can be understood as M-theory reductions on internal spaces with non-trivial topologies. It is worth mentioning that, by construction, all our models will admit a locally geometric description in the sense that they rely on an eleven-dimensional formulation correctly satisfying the section condition \cite{Berman:2011cg} in the language of Exceptional Field Theory (EFT) \cite{Hohm:2013vpa,Hohm:2013uia,Hohm:2014qga}.
This is in the spirit of ref.~\cite{Andriot:2012wx} and does not lead to non-geometric duality orbits in the sense of ref.~\cite{Dibitetto:2012rk}. However, such a formulation will in general only be equivalent to the traditional one up to total derivative terms \cite{Musaev:2014lna} that might play an important role upon reductions on non-toroidal topologies.

Even though our present work aims at shedding further light on the meaning of non-geometric fluxes, one cannot conclude anything about those non-geometric STU-models that were found to allow for stable dS critical points \cite{deCarlos:2009qm,Danielsson:2012by,Blaback:2013ht}. Whether it is possible to find novel examples of stable dS vacua satisfying the section condition still remains to be seen. Even so we expect that there will be compactness issues due to the no-go result in ref.~\cite{Maldacena:2000mw}.

The paper is organised as follows. We first present a group-theoretical truncation of maximal supergravity in four dimensions leading to isotropic STU-models with three complex scalars. We then employ some group theory arguments applied to the embedding tensor formalism in order to derive the flux-induced superpotentials describing M-theory compactifications on a twisted $T^{7}$, $S^{7}$ and $S^{4} \times T^{3}$. The result of this procedure will be a quadratic, quartic and cubic superpotential, respectively. We then discuss our results as well as some possible implications and future research directions. Finally, we collect some technical details concerning group theory in appendix~\ref{sec:app1}.

\section{M-theory on Different Geometries and Topologies}

The low-energy M-theory action in its democratic formulation reads
\be
\label{action11d}
S = \frac{1}{2\kappa_{11}^{2}} \, \int d^{11}x \, \sqrt{-g^{(11)}} \, \left(\mathcal{R}^{(11)} - \dfrac{1}{2} |G_{(4)}|^{2}  -  \dfrac{1}{2} |G_{(7)}|^{2}\right)  - \dfrac{1}{6 } \, \int   C_{(3)} 
\wedge G_{(4)}  \wedge G_{(4)} \ ,
\ee
where $|G_{(4)}|^{2} \ \equiv \ \frac{1}{4!} \, G_{(4) M_{1}\cdots M_{4}} \, {G_{(4)}}^{M_{1}\cdots M_{4}}$ and  
$|G_{(7)}|^{2} \ \equiv \ \frac{1}{7!} \, G_{(7) M_{1}\cdots M_{7}} \, {G_{(7)}}^{M_{1}\cdots M_{7}}$ with ${M=0, ..., 10}$.   
We choose the following reduction \emph{Ansatz}
\be
\label{Gen_red_ansatz}
ds^{2}_{(11)} \, = \, \tau^{-2} \, ds^{2}_{(4)} \, + \, \rho \, ds^{2}_{(7)} \ ,
\ee
where $\rho$ represents the volume of the internal space $X_{7}$ and $\tau$ is suitably determined,
\be
\tau \, = \, \rho^{7/4} \ ,
\ee
such that the \emph{Ansatz} in \eqref{Gen_red_ansatz} yield a 4D Lagrangian in the Einstein frame.

In the second part of this section we will be considering different choices for $X_{7}$ within the class of $S^{d} \times T^{7-d}$ leading to STU-models within $\mathcal{N}=1$ supergravity in 4D.
We will start out revisiting the case of a twisted $T^{7}$, and we will derive the flux-induced superpotential for this class of compactifications through group-theoretical considerations. This will help us
construct our working conventions, which will be used in the analogous derivations carried out for different choices of $X_{7}$ other than twisted tori.
Before we do this, we first need to introduce a particular group-theoretical truncation of maximal supergravity in 4D leading to the isotropic STU-models that we are interested in.

\subsection{An $\textrm{SO}(3) \times \mathbb{Z}_{2}$ truncation of $\mathcal{N}=8$ supergravity}

Maximal supergravity in 4D \cite{deWit:2007mt} enjoys $\textrm{E}_{7(7)}$ global symmetry and all its fields and deformations (\emph{i.e.} gaugings) transform into irrep's of such a global symmetry group. Vector fields transform in the $\textbf{56}$ though only half of them are physically independent due to electromagnetic duality, while scalar fields transform in the $\textbf{133}$, though
only $70$ of them are physically propagating due to the presence of a local $\textrm{SU}(8)$ symmetry. A group-theoretical truncation consists in branching all fields and deformations of the theory
into irrep's of a suitable subgroup $G_{0} \subset \textrm{E}_{7(7)}$ and retaining only the $G_{0}$-singlets. Such a truncation is guaranteed to be mathematically consistent due the $\textrm{E}_{7(7)}$
covariance of the eom's of maximal supergravity.

A first discrete $\mathbb{Z}_{2}$ truncation reads
\be
\begin{array}{ccl}
\textrm{E}_{7(7)} & \supset & \textrm{SL}(2)_{S} \times \textrm{SO}(6,6) \ , \\[2mm]
\textbf{56} & \overset{\mathbb{Z}_{2}}{\rightarrow} & (\textbf{2},\textbf{12})_{(+)} \oplus (\textbf{1},\textbf{32})_{(-)} \ ,
\end{array}
\nonumber
\ee
where only the $\mathbb{Z}_{2}$-even irrep's are retained in the truncation\footnote{From a more physical perspective, such a $\mathbb{Z}_{2}$ can be understood as an orientifold involution for those
supergravities coming from reductions of type II theories.}. 
This procedure yields (gauged) $\mathcal{N}=4$ supergravity in $D=4$ \cite{Dibitetto:2011eu}.

In the second step, we perform a truncation to the $\textrm{SO}(3)$-invariant sector in the following way
\be
\label{chain}
\textrm{SL}(2)_{S} \times \textrm{SO}(6,6) \,\,\supset \,\,\textrm{SL}(2)_{S} \times \textrm{SO}(2,2) \times \textrm{SO}(3) \,\sim\,  \prod_{\Phi=S,T,U} \textrm{SL}(2)_{\Phi} \,\times\, \textrm{SO}(3) \ .
\ee
This step breaks half-maximal to minimal $\,\mathcal{N}=1\,$ supergravity due to the decomposition $\,\textbf{4} \rightarrow \textbf{1} \oplus \textbf{3}\,$ of the fundamental representation of the 
$\,\textrm{SU}(4)\,$ $R$-symmetry group in $\mathcal{N}=4$ supergravity under the $\,\textrm{SO}(3)\,$ subgroup
\be
\textrm{SU}(4) \,\,\supset \,\, \textrm{SU}(3) \,\,\supset \,\, \textrm{SO}(3) \ .
\ee
The resulting theory does not contain vectors since there are no $\,\textrm{SO}(3)$-singlets in the decomposition $\,\textbf{12} \rightarrow (\textbf{4},\textbf{3})\,$ of the fundamental representation of 
$\,\textrm{SO}(6,6)\,$ under $\,\textrm{SO}(2,2) \times \textrm{SO}(3)$. The physical scalar fields span the coset space
\be
\mathcal{M}_{\textrm{scalar}}= \prod_{\Phi=S,T,U}\left( \frac{\textrm{SL}(2)}{\textrm{SO}(2)} \right)_{\Phi} \ ,
\ee
involving three $\,\textrm{SL}(2)/\textrm{SO}(2)\,$ factors each of which can be parameterised by the complex scalars $\,\Phi=(S , T , U)$. Such scalars can be obtained by decomposing the adjoint representation $\textbf{133}$
of $\textrm{E}_{7(7)}$ according to the chain in \eqref{chain} to find \emph{nine} real $\textrm{SO}(3)$-singlets, out of which only six correspond to physical dof's.

The K\"ahler potential of the theory reads
\be
K \ = \ -\log\left(-i\,(S-\bar{S})\right) \, - \, 3\log\left(-i\,(T-\bar{T})\right) \, - \, 3\log\left(-i\,(U-\bar{U})\right) \ .
\ee
In addition, the embedding tensor of the theory contains $\,40\,$ independent components (coming this time from the decomposition of the $\textbf{912}$ of $\textrm{E}_{7(7)}$ according to the chain in \eqref{chain}) which can be viewed as the superpotential couplings\footnote{The connection between the $\,\mathcal{N}=1\,$ and
$\mathcal{N}=4$ theory was extensively investigated in refs~\cite{Aldazabal:2008zza,Dall'Agata:2009gv}. However, the explicit agreement between the  scalar potentials up to quadratic constraints was first shown in 
ref.~\cite{Dibitetto:2011gm}.} representing a complete duality-inviariant set of \emph{generalised fluxes} \cite{Shelton:2005cf}. 
This yields the following duality-covariant flux-induced superpotential 
\be
\label{W_fluxes4}
\mathcal{W} \ = \ (P_{F} - P_{H} \, S ) + 3 \, T \, (P_{Q} - P_{P} \, S ) + 3 \, T^2 \, (P_{Q'} - P_{P'} \, S ) + T^3 \, (P_{F'} - P_{H'} \, S ) \ ,
\ee
involving the three complex moduli $S$, $T$ and $U$ surviving the SO($3$)-truncation introduced ealier in this section. 
\be
\begin{array}{lcll}
\label{Poly_unprim}
P_{F} = a_0 - 3 \, a_1 \, U + 3 \, a_2 \, U^2 - a_3 \, U^3 & \hspace{5mm},\hspace{5mm} & P_{H} = b_0 - 3 \, b_1 \, U + 3 \, b_2 \, U^2 - b_3 \, U^3 & ,  \\[2mm]
P_{Q} = c_0 + C_{1} \, U - C_{2} \, U^2 - c_3 \, U^3 & \hspace{5mm},\hspace{5mm} & P_{P} = d_0 + D_{1} \, U - D_{2} \, U^2 - d_3 \, U^3 & ,
\end{array}
\ee
as well as those induced by their primed counterparts $\,(F',H')\,$ and $\,(Q',P')\,$ fluxes \cite{Aldazabal:2006up},
\be
\begin{array}{lcll}
\label{Poly_prim}
P_{F'} = a_3' + 3 \, a_2' \, U + 3 \, a_1' \, U^2 + a_0' \, U^3 & \hspace{3mm},\hspace{3mm} &P_{H'} = b_3' + 3 \, b_2' \, U + 3 \, b_1' \, U^2 + b_0' \, U^3 & ,  \\[2mm]
P_{Q'} = -c_3' +  C'_{2} \, U + C'_{1} \, U^2 - c_0' \, U^3 & \hspace{3mm},\hspace{3mm} & P_{P'} = -d_3' + D'_{2} \, U + D'_{1} \, U^2 - d_0' \, U^3 & .
\end{array}
\ee
For the sake of simplicity, we have introduced the flux combinations $\,C_{i} \equiv 2 \, c_i - \tilde{c}_{i}\,$, $\,D_{i} \equiv 2 \, d_i - \tilde{d}_{i}\,$, $\,C'_{i} \equiv 2 
\, c'_i - \tilde{c}'_{i}\,$ and $\,D'_{i} \equiv 2 \, d'_i - \tilde{d}'_{i}\,$ entering the superpotential (\ref{W_fluxes4}), and hence also the scalar potential.

In order to relate our 4D deformed supergravity models to M-theory reductions on different geometries and topologies, one needs to fix some conventions for assigning a $\mathbb{Z}_{2}$ parity to the seven
physical coordinates on $X_{7}$. We adopt a set of conventions that is inherited from the link with type IIA compactifications with $\textrm{O}6$-planes \cite{Dibitetto:2014sfa}, where such a parity 
transformation can be viewed as orientifold involution. 
\be
\begin{array}{lclclclc}
x^{M} & \longrightarrow & \underbrace{x^{\mu}}_{\textrm{4D}} & \oplus & \underbrace{x^{a}}_{(+)} & \oplus & \underbrace{x^{i} \,\oplus\, x^{7}}_{(-)} & ,
\end{array}\label{parity}
\ee
where $x^{m}\,\equiv\,\left(x^{a}, \, x^{i}, \, x^{7}\right)$ realise the compact geometry of $X_{7}$. Retaining only even fields and fluxes w.r.t. the action of the above $\mathbb{Z}_{2}$ will automatically restrict
the supergravity theory obtained through an M-theory reduction to the framework of $\mathcal{N}=1$ STU-models.   

The metric \eqref{Gen_red_ansatz} splits accordingly into
\be
\label{M_red_ansatz}
ds^{2}_{11} \, = \, \rho^{-7/2} \, ds^{2}_{4} \, + \, \rho \, \left(\sigma^{-1} \kappa^{-1}\, M_{ab} \, \eta^{a} \otimes \eta^{b} \, + \, \sigma \kappa^{-1} \, M_{ij} \, 
\eta^{i} \otimes \eta^{j} \, + \, \kappa^{6} \left(\eta^{7}\right)^{2}\right) \ ,
\ee
where $\left\{\eta^{m}\right\}\,\equiv\,\left\{\eta^{a}, \, \eta^{i}, \, \eta^{7}\right\}$ represents a basis of one-forms carrying information about the dependence of the metric on the internal coordinates.  
 The $\mathbb{R}^{+}$ scalars $\sigma$ and $\kappa$ parametrise the relative size between the $a$ and $i$ coordinates, which acquire opposite involution-parity when adopting the type IIA picture 
\cite{Danielsson:2012et} and the relative size between the type IIA directions and the M-theory circle, respectively. 
Moreover, $M_{ab}$ and $M_{ij}$ contain in general $\textrm{SL}(3)_{a} \, \times \, \textrm{SL}(3)_{i}$ scalar excitations. However, such degrees of freedom are frozen due to the requirement of
 $\textrm{SO}(3)$-invariance, \emph{i.e.} $M_{ab} \, = \, \delta_{ab}$ and $M_{ij} \, = \, \delta_{ij}$. 

The relationship between the STU scalars and the above geometric moduli reads
\be
\label{STU2rhokappasigma}
\begin{array}{lclclc}
\textrm{Im}(S) \ = \ \rho^{3/2} \, \left(\frac{\kappa}{\sigma}\right)^{3/2} & , & \textrm{Im}(T) \ = \ \rho^{3/2} \,\frac{\sigma^{1/2}}{\kappa^{3/2}} & , & 
\textrm{Im}(U) \ = \ \rho^{3/2} \, \kappa^{2}  & .
\end{array}
\ee

\subsection{Compactifications on a twisted $T^7$}

The seven compact coordinates of the torus transform in the fundamental representation of the $\textrm{SL}(7)$ subgroup of $\textrm{E}_{7(7)}$, which can be viewed as the group of diffeomorphisms on $T^{7}$
with twist. The relevant chain of decompositions is
\be
\begin{array}{lclclclc}
\textrm{E}_{7(7)}  & \supset  & \textrm{SL}(8) & \supset  & \mathbb{R}^{+}_{M} \times \textrm{SL}(7) & \supset  & \mathbb{R}^{+}_{M} \times \mathbb{R}^{+}_{B} \times \textrm{SL}(6)  & , 
\end{array}\notag
\ee
and finally down to 
\be
\mathbb{R}^{+}_{M} \times \mathbb{R}^{+}_{B} \times \mathbb{R}^{+}_{A} \times \textrm{SL}(3)_{a} \times \textrm{SL}(3)_{i} \ , 
\notag
\ee
where one should, furthermore, only restrict to \emph{isotropic} objects. Following the philosophy of ref.~\cite{Dibitetto:2014sfa}, one can match the na\"ive scaling behaviour coming from dimensional 
reductions of the various terms in the action \eqref{action11d} with the correct STU-charges by using the relations in \eqref{STU2rhokappasigma}. This results in the following mapping 
\be
\left\{
\begin{array}{lclclclc}
q_{S}  & =  & \frac{1}{28} q_{M} \, - \, \frac{1}{28} q_{B} \, - \, \frac{1}{4} q_{A} & , \\[2mm]
q_{T}  & =  & \frac{3}{28} q_{M} \, - \, \frac{3}{28} q_{B} \, + \, \frac{1}{4} q_{A} & , \\[2mm]
q_{U}  & =  & \frac{3}{28} q_{M} \, + \, \frac{1}{7} q_{B}   & , 
\end{array}\right.
\label{MBA/STU}
\ee
between the group-theoretical $\mathbb{R}^{+}$ charges obtained from the above decomposition and the STU-charges realised in $\mathcal{N}=1$ supergravity.
Such a mapping allows one to derive a dictionary between fluxes and superpotential couplings. 

\begin{table}[t!]
\begin{center}
\begin{tabular}{|c|c|c|c|}
\hline
STU couplings & M-theory fluxes & Flux labels & $\mathbb{R}^{+}_{S} \times \mathbb{R}^{+}_{T} \times \mathbb{R}^{+}_{U} \times \textrm{SL}(3)_{a} \times \textrm{SL}(3)_{i}$ irrep's \\
\hline
\hline
$1$ & $G_{aibjck7}$ & $a_{0}$ & $(\textbf{1},\textbf{1})_{(+\frac{1}{2};+\frac{3}{2};+\frac{3}{2})}$ \\
\hline
\hline
$S$ & $G_{ijk7}$ & $b_{0}$ & $(\textbf{1},\textbf{1})_{(-\frac{1}{2};+\frac{3}{2};+\frac{3}{2})}$\\
\hline
$T$ & $G_{ibc7}$ & $c_{0}$ & $(\textbf{3},\textbf{3}^{\prime})_{(+\frac{1}{2};+\frac{1}{2};+\frac{3}{2})}$ \\
\hline
$U$ & $G_{aibj}$ & $a_{1}$ & $(\textbf{3},\textbf{3})_{(+\frac{1}{2};+\frac{3}{2};+\frac{1}{2})}$\\
\hline
\hline
$S \, T$ & ${\omega_{7i}}^{a}$ & $d_{0}$ & $(\textbf{3},\textbf{3}^{\prime})_{(-\frac{1}{2};+\frac{1}{2};+\frac{3}{2})}$\\
\hline
$T^{2}$ & ${\omega_{a7}}^{i}$ &  $c_{3}^{\prime}$ & $(\textbf{3}^{\prime},\textbf{3})_{(+\frac{1}{2};-\frac{1}{2};+\frac{3}{2})}$\\
\hline
$T \, U$  & ${\omega_{a j}}^{k}$, ${\omega_{b c}}^{a}$  & $c_{1}$, $\tilde{c}_{1}$  & $\left((\textbf{3}^{\prime},\textbf{8})\oplus(\textbf{6},\textbf{1})\right) _{(+\frac{1}{2};+\frac{1}{2};+\frac{1}{2})}$\\
\hline
$S \, U$ & ${\omega_{jk}}^{a}$ & $ b_{1}$ & $(\textbf{3},\textbf{3})_{(-\frac{1}{2};+\frac{3}{2};+\frac{1}{2})}$ \\
\hline
$U^{2} $ & ${\omega_{ai}}^{7}$ & $a_{2}$ & $(\textbf{3}^{\prime},\textbf{3}^{\prime})_{(+\frac{1}{2};+\frac{3}{2};-\frac{1}{2})}$ \\
\hline
\end{tabular}
\end{center}
\caption{{\it Summary of M-theory fluxes and superpotential couplings on a twisted $T^{7}$. Isotropy (\emph{i.e.} $\textrm{SO}(3)$-invariance) only allows for flux components that can be constructed by 
using $\epsilon_{(3)}$'s and $\delta_{(3)}$'s. 
These symmetries also induce a natural splitting $\,\eta^{A}=(\eta^{a} \,,\, \eta^{i} \,,\, \eta^{7})\,$ where $\,a=1,3,5\,$ and $\,i=2,4,6\,$.}}
\label{Table:Fluxes_T7}
\end{table}

From the decomposition of the fundamental representation of $\textrm{E}_{7(7)}$ (see appendix~\ref{sec:app1} for the details)
\be
\begin{array}{cclc}
\textrm{E}_{7(7)}  & \supset  & \mathbb{R}^{+}_{M} \times \mathbb{R}^{+}_{B} \times \mathbb{R}^{+}_{A} \times \textrm{SL}(3)_{a} \times \textrm{SL}(3)_{i}  & , \\[3mm]
\textbf{56}  & \rightarrow  & (\textbf{1},\textbf{1})_{(+6;+6;0)} \oplus (\textbf{3}^{\prime},\textbf{1})_{(+6;-1;-1)} \oplus 
(\textbf{1},\textbf{3}^{\prime})_{(+6;-1;+1)} \oplus \, \dots & ,
\end{array}
\ee
one can exactly and unambiguously identify the physical derivative operators along the seven M-theory internal directions as
\be
\begin{array}{lclcl}
\partial_{a} \, \in \,(\textbf{3}^{\prime},\textbf{1})_{(+\frac{1}{2};+\frac{1}{2};+\frac{1}{2})} & ,  & \partial_{i} \, \in \,(\textbf{1},\textbf{3}^{\prime})_{(0;+1;+\frac{1}{2})} & , &
\partial_{7} \, \in \,(\textbf{1},\textbf{1})_{(0;0;+\frac{3}{2})} 
\end{array}\label{derivatives}
\ee
w.r.t. $\mathbb{R}^{+}_{S} \times \mathbb{R}^{+}_{T} \times \mathbb{R}^{+}_{U} \times \textrm{SL}(3)_{a} \times \textrm{SL}(3)_{i}$.

As far as the fluxes are concerned, one has to decompose the embedding tensor of maximal supergravity 
\be
\begin{array}{cclc}
\textrm{E}_{7(7)}  & \supset  & \textrm{SL}(8)  & , \\[3mm]
\textbf{912}  & \rightarrow  & \textbf{36} \oplus \textbf{36}^{\prime} \oplus \textbf{420} \oplus \textbf{420}^{\prime} & ,
\end{array}
\ee
to find
\be
\begin{array}{cclc}
\textrm{SL}(8)  & \supset  & \mathbb{R}^{+}_{M} \times \mathbb{R}^{+}_{B} \times \mathbb{R}^{+}_{A} \times \textrm{SL}(3)_{a} \times \textrm{SL}(3)_{i}  & , \\[3mm]
\textbf{36}^{\prime}  & \rightarrow  & (\textbf{1},\textbf{1})_{(+14;0;0)} \oplus \, \dots & ,\\[2mm]
\textbf{420}  & \rightarrow  & (\textbf{1},\textbf{1})_{(+10;+3;+3)} \oplus (\textbf{3},\textbf{3})_{(+10;-4;0)} \oplus (\textbf{3},\textbf{3}^{\prime})_{(+10;+3;-1)} \oplus \, \dots & ,\\[2mm]
\textbf{420}^{\prime}  & \rightarrow  & (\textbf{3},\textbf{3})_{(+6;-1;+3)} \oplus (\textbf{6},\textbf{1})_{(+6;-1;-1)} \oplus (\textbf{3}^{\prime},\textbf{8})_{(+6;-1;-1)} \oplus \\
 &  &  (\textbf{3}^{\prime},\textbf{3}^{\prime})_{(+6;-8;0)} \oplus (\textbf{3}^{\prime},\textbf{3})_{(+6;+6;-2)} \oplus (\textbf{3},\textbf{3}^{\prime})_{(+6;+6;+2)} \oplus \, \dots & ,
\end{array}
\ee
where we have used the decomposition in \eqref{E7->SL8}. By using the dictionary \eqref{MBA/STU}, we were able to reproduce all the correct STU scalings of the fluxes on a twisted $T^{7}$.
The results of this procedure are collected in table~\ref{Table:Fluxes_T7} and agree with those already found earlier in refs~\cite{Dall'Agata:2005fm, Derendinger:2014wwa}. 
As a consequence, the flux-induced superpotential in this case reads
\be
\label{W_T7}
\mathcal{W}_{(T^{7})} \ = \ a_{0} \, - \, b_{0}S \, + \, 3c_{0}T \, - \, 3a_{1}U \, + \,3 a_{2}U^{2} \, + \, 3(2c_{1}-\tilde{c}_{1})TU \, + \, 3b_{1} SU \, - \, 3c_{3}^{\prime}T^{2} \, - \, 3d_{0}ST \ .
\ee
One should note that the underlying gauging for this class of compactifications is expected to be non-semisimple, its semisimple part being the group realised by the components of $\omega$-flux as structure 
constants. The non-semisimple extension is given by the presence of 4- and 7-form gauge fluxes.
This is in line with what already observed in refs~\cite{Dall'Agata:2009gv,Dibitetto:2012ia,Dibitetto:2014sfa} in the context of massive type IIA compactifications on a twisted $T^{6}$ in the absence of local sources, 
where the corresponding effective 4D description turned out to be $\mathcal{N}=8$ supergravity with gauge group $\textrm{SO}(4) \ltimes \textrm{Nil}_{22}$.

\subsection{Compactifications on $S^7$}

Let us now consider the compactification of M-theory on $S^7$. In refs~\cite{Freund:1980xh,Englert:1982vs} it was already noted that such a compactification is described by an $\textrm{SO}(8)$ gauging in 4D maximal supergravity.
The components of the embedding tensor are parametrised by a symmetric $8 \times 8$ matrix $\Theta_{AB}$ transforming\footnote{We adopt the following conventions $X^{A} \,\equiv\, \left(X^{a},\,X^{i},\,
X^{7},\,X^{8}\right)$ for the fundamental representation of $\textrm{SL}(8)$.} in the $\textbf{36}^{\prime}$ of $\textrm{SL}(8)$. 

Gaugings in the $\textbf{36}\,\oplus\,\textbf{36}^{\prime}$ are in general identified by $\tilde{\Theta}^{AB}\,\oplus\,\Theta_{AB}$ satisfying the following \emph{Quadratic Constraints} \cite{DallAgata:2011aa}
\be
\label{QC}
\Theta_{AC} \, \tilde{\Theta}^{CB} \ - \ \frac{1}{8} \, \left(\Theta_{CD} \, \tilde{\Theta}^{CD}\right) \, \delta^{B}_{A} \ = \ 0 \ .
\ee
Such theories have a subgroup of $\textrm{SL}(8)$ as gauge group and their scalar potential can be written in terms of a complex pseudo-superpotential \cite{Borghese:2012zs} 
\be
\label{VfromW}
V \, = \, -\frac{3}{8} \, |W|^{2} \, + \, \frac{1}{4} \, |\partial W|^{2} \ ,
\ee
where $W \, \equiv \, \frac{1}{2} \, \left(\Theta_{AB}\mathcal{M}^{AB} \, - \, i \,\tilde{\Theta}^{AB}\mathcal{M}_{AB}\right)$, $\mathcal{M}^{AB}$ being the $\textrm{SL}(8)/\textrm{SO}(8)$ coset
representative and $\mathcal{M}_{AB}$ its inverse. 

In the relevant $S^7$ example, the embedding tensor reads
\be
\begin{array}{lclc}
\Theta_{AB} \ = \ \left(\begin{array}{c|c|c|c}-\tilde{c}_{1}^{\prime} \,\mathds{1}_{3} & & & \\[2mm]\hline & -\tilde{d}_{2} \,\mathds{1}_{3} & & \\[2mm]\hline & & -b_{3}^{\prime} & \\[2mm]\hline
 & & & a_{0}\end{array}\right) \ = \ \mathds{1}_{8} & , & \tilde{\Theta}^{AB} \ = \ 0_{8} & ,
\end{array}
\ee
which belongs to the \emph{semisimple} branch of solutions to the constraints in \eqref{QC}. The corresponding scalar potential in \eqref{VfromW} simplifies to 
\be
\label{VN=8}
V \, = \, \frac{1}{8} \, \Theta_{AB}\Theta_{CD}\left(2\,\mathcal{M}^{AC}\mathcal{M}^{BD} \, - \, \mathcal{M}^{AB}\mathcal{M}^{CD}\right) \ .
\ee

We will now interpret this theory as an STU-model and rederive the corresponding flux-induced superpotential by means of group theory arguments.
In this case the relevant decomposition is still the same as the one in the twisted $T^{7}$ case
\be
\hspace{-4.8mm}
\begin{array}{cclc}
\textrm{SL}(8)  & \supset  & \mathbb{R}^{+}_{M} \times \mathbb{R}^{+}_{B} \times \mathbb{R}^{+}_{A} \times \textrm{SL}(3)_{a} \times \textrm{SL}(3)_{i}  & , \\[3mm]
\textbf{36}^{\prime}  & \rightarrow  & (\textbf{1},\textbf{1})_{(+14;0;0)} \oplus (\textbf{1},\textbf{1})_{(-2;+12;0)} \oplus (\textbf{6}^{\prime},\textbf{1})_{(-2;-2;-2)} \oplus 
(\textbf{1},\textbf{6})_{(-2;-2;+2)} \oplus \, \dots & .
\end{array}
\ee
This gives the STU-couplings collected in table~\ref{Table:Fluxes_S7} upon using the dictionary \eqref{MBA/STU}.
\begin{table}[t!]
\begin{center}
\begin{tabular}{|c|c|c|c|}
\hline
STU couplings & M-theory fluxes & Flux labels & $\mathbb{R}^{+}_{S} \times \mathbb{R}^{+}_{T} \times \mathbb{R}^{+}_{U} \times \textrm{SL}(3)_{a} \times \textrm{SL}(3)_{i}$ irrep's \\
\hline
\hline
$1$ & $G_{aibjck7}$ & $a_{0}$ & $(\textbf{1},\textbf{1})_{(+\frac{1}{2};+\frac{3}{2};+\frac{3}{2})}$ \\
\hline
\hline
$S\,T^{3}$ & $\Theta_{77}$ &  $b_{3}^{\prime}$ & $(\textbf{1},\textbf{1})_{(-\frac{1}{2};-\frac{3}{2};+\frac{3}{2})}$\\
\hline
$T^{2} \, U^{2}$ & $\Theta_{ab}$ & $ \tilde{c}_{1}^{\prime}$ & $(\textbf{6}^{\prime},\textbf{1})_{(+\frac{1}{2};-\frac{1}{2};-\frac{1}{2})}$ \\
\hline
$S \, T \, U^{2} $ & $\Theta_{ij}$ & $\tilde{d}_{2}$ & $(\textbf{1},\textbf{6}^{\prime})_{(-\frac{1}{2};+\frac{1}{2};-\frac{1}{2})}$ \\
\hline
\hline
$T \, U $ & $\Theta_{i7}$ & $c_{1}$ & $(\textbf{3}^{\prime},\textbf{1})_{(+\frac{1}{2};+\frac{1}{2};+\frac{1}{2})}$ \hspace{4mm} (non-isotropic) \\
\hline
\end{tabular}
\end{center}
\caption{{\it Summary of M-theory fluxes and superpotential couplings on $S^{7}$. Isotropy (\emph{i.e.} $\textrm{SO}(3)$-invariance) only allows for flux components that can be constructed by using $\epsilon_{(3)}$'s
and $\delta_{(3)}$'s. In the frame we have chosen one of the objects in the $\textbf{36}^{\prime}$ is $G_{(7)}$ flux, whereas the quartic couplings describe the $S^{7}$ geometry.}}
\label{Table:Fluxes_S7}
\end{table}
The associated flux-induced superpotential is given by
\be
\label{W_S7}
\mathcal{W}_{(S^{7})} \ = \ a_{0} \, - \, b_{3}^{\prime}ST^{3} \, - \, 3\tilde{c}_{1}^{\prime}T^{2}U^{2} \, - \, 3 \tilde{d}_{2}STU^{2} \ ,
\ee
which matches what was found in refs~\cite{Dibitetto:2012ia, Dibitetto:2012xd} in the context of STU-models. The $\mathcal{N}=1$ scalar potential computed from \eqref{W_S7}
coincides with \eqref{VN=8} upon using the correct identification of the STU scalars inside the coset representative $\mathcal{M}^{AB}$.

\subsection{Compactifications on $S^4 \times T^3$}

We have seen how for $S^7$ the superpotential contains only the constant part and some quartic parts. We will now analyse the flux-induced superpotential for $S^4 \times T^3$ to find that
cubic terms will appear, thus mimicking the effect of the presence of $Q$-flux.

Given the natural factorisation that $X_{7}$ has in this case, the relevant branching one should analyse goes through
\be
\begin{array}{lclclclc}
\textrm{E}_{7(7)}  & \supset  & \textrm{SL}(8) & \supset  & \mathbb{R}^{+}_{Q} \times \textrm{SL}(3)_{a} \times \textrm{SL}(5) & \supset  & 
\mathbb{R}^{+}_{Q} \times \mathbb{R}^{+}_{1} \times \textrm{SL}(3)_{a} \times \textrm{SL}(4)  & , 
\end{array}\notag
\ee
and finally down to 
\be
\mathbb{R}^{+}_{Q} \times \mathbb{R}^{+}_{1} \times \mathbb{R}^{+}_{2} \times \textrm{SL}(3)_{a} \times \textrm{SL}(3)_{i} \ , 
\notag
\ee
where, as usual, only isotropic objects should be retained within our STU-model.

By following the new branching of the fundamental representation of $\textrm{E}_{7(7)}$ 
\be
\begin{array}{cclc}
\textrm{E}_{7(7)}  & \supset  & \mathbb{R}^{+}_{Q} \times \mathbb{R}^{+}_{1} \times \mathbb{R}^{+}_{2} \times \textrm{SL}(3)_{a} \times \textrm{SL}(3)_{i}  & , \\[3mm]
\textbf{56}  & \rightarrow  & (\textbf{1},\textbf{1})_{(-6;+3;+3)} \oplus (\textbf{3}^{\prime},\textbf{1})_{(-10;0;0)} \oplus 
(\textbf{1},\textbf{3}^{\prime})_{(-6;+3;-1)} \oplus \, \dots & ,
\end{array}
\ee
and demanding that the physical derivative operators identified in \eqref{derivatives} be the same, one finds 
\be
\left\{
\begin{array}{lclclclc}
q_{S}  & =  & -\frac{1}{20} q_{Q} \, - \, \frac{1}{10} q_{1}  & , \\[2mm]
q_{T}  & =  & -\frac{1}{20} q_{Q} \, + \, \frac{3}{20} q_{1} \, - \, \frac{1}{4} q_{2} & , \\[2mm]
q_{U}  & =  & -\frac{1}{20} q_{Q} \, + \, \frac{3}{20} q_{1}  \, + \, \frac{1}{4} q_{2} & , 
\end{array}\right.
\label{Q12/STU}
\ee
as a new dictionary between STU-scaling weights and group-theoretical $\mathbb{R}^{+}_{Q} \times \mathbb{R}^{+}_{1} \times \mathbb{R}^{+}_{2}$ charges. Note that this procedure of identifying the
seven physical derivative operators corresponds to choosing the relevant solution to the \emph{section condition} in the EFT sense. In this case, several 
\emph{a priori} different choices are possible but they all yield a superpotential that is unique up to modular transformations.

As far as the fluxes are concerned, in total analogy with the $S^{7}$ case, now we expect to be able to describe the $S^{4}$ geometry with $G_{(4)}$ flux by turning on embedding tensor deformations transforming 
in the $\textbf{15}^{\prime}$ of $\textrm{SL}(5)$, \emph{i.e.} a symmetric $5 \times 5$ matrix $\Theta_{IJ}$. This would in itself lead to a maximal $\textrm{SO}(5)$-gauged supergravity in 7D 
\cite{Samtleben:2005bp}. 

However, these deformations can be supplemented with $G_{(7)}$ flux wrapping the whole $X_{7}$ and a twisting on the $T^{3}$ producing some metric flux $\omega$. Due to the different $\mathbb{Z}_{2}$-parity
assigned to the M-theory coordinates through \eqref{parity}, there are \emph{three} inequivalent models that one can study, each of them characterised by different flux components threading internal
space:
\be
\begin{array}{ccccc}
{\bf Model \ 1:} & & {\bf Model \ 2:} & & {\bf Model \ 3:} \\[3mm]
\underbrace{\begin{array}{ccc}
\overset{a}{+} & \overset{b}{+} & \overset{c}{+} \\
\end{array}}_{T^{3}} \ 
\underbrace{\begin{array}{cccc}
\overset{i}{-} & \overset{j}{-} & \overset{k}{-} & \overset{7}{-} 
\end{array}}_{S^{4}} & &
\underbrace{\begin{array}{ccc}
\overset{a}{+} & \overset{j}{-} & \overset{k}{-} \\
\end{array}}_{T^{3}} \ 
\underbrace{\begin{array}{cccc}
\overset{b}{+} & \overset{c}{+} & \overset{i}{-} & \overset{7}{-} 
\end{array}}_{S^{4}} & &
\underbrace{\begin{array}{ccc}
\overset{a}{+} & \overset{i}{-} & \overset{7}{-} \\
\end{array}}_{T^{3}} \ 
\underbrace{\begin{array}{cccc}
\overset{b}{+} & \overset{c}{+} & \overset{j}{-} & \overset{k}{-} 
\end{array}}_{S^{4}}
\end{array}\notag
\ee
Out of these {\bf Model 1} is the only choice that is compatible with $\textrm{SO}(3)$-invariance thus yielding an \emph{isotropic} STU-model. This is the model we will focus on, and for which we will provide
details. Dealing with {\bf Model 2} and {\bf Model 3} requires further breaking $\textrm{SL}(3)_{a} \times \textrm{SL}(3)_{i}$ symmetry down to 
$\textrm{SL}(2)_{a} \times \textrm{SL}(2)_{i} \times \mathbb{R}^{+}_{a} \times \mathbb{R}^{+}_{i}$, this giving rise to \emph{non-isotropic} STU-models. We will only sketchily show that such
non-isotropic superpotentials will still be cubic.

\begin{itemize}

\item {\bf Model 1:} In this case our decomposition contains the following relevant pieces
\be
\begin{array}{cclc}
\textrm{SL}(8)  & \supset  & \mathbb{R}^{+}_{Q} \times \mathbb{R}^{+}_{1} \times \mathbb{R}^{+}_{2} \times \textrm{SL}(3)_{a} \times \textrm{SL}(3)_{i}  & , \\[3mm]
\textbf{36}  & \rightarrow  & (\textbf{6},\textbf{1})_{(-10;0;0)} \oplus \, \dots & ,\\[2mm]
\textbf{36}^{\prime}  & \rightarrow  & (\textbf{1},\textbf{1})_{(-6;+8;0)} \oplus (\textbf{1},\textbf{6}^{\prime})_{(-6;-2;-2)} \oplus (\textbf{1},\textbf{1})_{(-6;-2;+6)} \oplus \, \dots & ,\\[2mm]
\textbf{420}  & \rightarrow  & (\textbf{1},\textbf{1})_{(-18;+4;0)} \oplus \, \dots & ,
\end{array}
\ee
Using the relations in \eqref{Q12/STU}, we derived the fluxes activated by the $S^{4} \times T^{3}$ compactifications realised according to the first of the three different models presented above. The results
of this procedure are collected and shown in table~\ref{Table:Fluxes_S4T3_1}.
\begin{table}[t!]
\begin{center}
\begin{tabular}{|c|c|c|c|}
\hline
STU couplings & M-theory fluxes & Flux labels & $\mathbb{R}^{+}_{S} \times \mathbb{R}^{+}_{T} \times \mathbb{R}^{+}_{U} \times \textrm{SL}(3)_{a} \times \textrm{SL}(3)_{i}$ irrep's \\
\hline
\hline
$1$ & $G_{aibjck7}$ & $a_{0}$ & $(\textbf{1},\textbf{1})_{(+\frac{1}{2};+\frac{3}{2};+\frac{3}{2})}$ \\
\hline
\hline
$S$ & $G_{ijk7}$ & $b_{0}$ & $(\textbf{1},\textbf{1})_{(-\frac{1}{2};+\frac{3}{2};+\frac{3}{2})}$\\
\hline
\hline
$T \, U$  &  ${\omega_{b c}}^{a}$  &  $\tilde{c}_{1}$  & $(\textbf{6},\textbf{1})_{(+\frac{1}{2};+\frac{1}{2};+\frac{1}{2})}$\\
\hline
\hline
$T^{3}$ & $\Theta_{77}$ &  $a_{3}^{\prime}$ & $(\textbf{1},\textbf{1})_{(+\frac{1}{2};-\frac{3}{2};+\frac{3}{2})}$\\
\hline
$T \, U^{2} $ & $\Theta_{ij}$ & $\tilde{c}_{2}$ & $(\textbf{1},\textbf{6}^{\prime})_{(+\frac{1}{2};+\frac{1}{2};-\frac{1}{2})}$ \\
\hline
\hline
$T^{2} \, U $ & $\Theta_{i7}$ & $\tilde{c}_{2}^{\prime}$ & $(\textbf{1},\textbf{3}^{\prime})_{(+\frac{1}{2};-\frac{1}{2};+\frac{1}{2})}$ \hspace{4mm} (non-isotropic) \\
\hline
\end{tabular}
\end{center}
\caption{{\it Summary of M-theory fluxes and superpotential couplings on a twisted $S^{4}\times T^{3}$ according to Model 1. 
Isotropy (\emph{i.e.} $\textrm{SO}(3)$-invariance) only allows for flux components that can be constructed by using $\epsilon_{(3)}$'s and $\delta_{(3)}$'s. 
Our chosen frame includes $G_{(4)}$ flux as one of the objects sitting in the $\textbf{36}^{\prime}$, whereas the other ones there parametrise the $S^{4}$ geometry.}}
\label{Table:Fluxes_S4T3_1}
\end{table}
The associated flux-induced superpotential is given by
\be
\label{W_S4T3}
\mathcal{W}_{(S^{4} \times T^{3})} \ = \ a_{0} \, - \, b_{0}S  \, - \, 3\tilde{c}_{1}TU \, + \, a_{3}^{\prime}T^{3} \, - \, 3 \tilde{c}_{2}TU^{2} \ ,
\ee
which contains some cubic contributions that can be regarded as M-theory $Q$-flux ${Q_{A}}^{[BCD]}$. The explicit relation between $Q$-flux and the components of the $\Theta$ tensor reads \cite{Blair:2014zba}
\be
\Theta_{AB} \ = \ \frac{1}{3!} \, {Q_{(A}}^{CDE} \, \epsilon_{B)CDE} \ ,
\notag
\ee
where $A,\,B,\,\dots\,=\,i,\,7$ represents a fundamental index on $S^{4}$. 

One should note that the geometry of the twisted $T^{3}$ sits in the $\textbf{36}$. The explicit way the corrsponding $\omega$-flux is embedded in $\tilde{\Theta}$ is given by
\be
\tilde{\Theta}^{ab} \ = \ \frac{1}{2!} \, {\omega_{cd}}^{(a} \, \epsilon^{b)cd} \ ,
\ee
where now the indices $a,\,b,\,c,\,\dots$ label the legs of the $T^{3}$.

Thus, in contrast with the $S^{7}$ case, such a background lies in the \emph{non-semisimple} branch of solutions to the constraints \eqref{QC}
\be
\begin{array}{lclc}
\Theta_{AB} \ = \ \left(\begin{array}{c|c|c|c}0_{3} & & & \\[2mm]\hline & \tilde{c}_{2} \,\mathds{1}_{3} & & \\[2mm]\hline & & a_{3}^{\prime} & \\[2mm]\hline
 & & & b_{0}\end{array}\right) \ = \ \left(\begin{array}{c|c}0_{3} &  \\[2mm]\hline & \mathds{1}_{5} \end{array}\right) & , & 
\tilde{\Theta}^{AB} \ = \ \left(\begin{array}{c|c} \tilde{c}_{1} \,\mathds{1}_{3} &  \\[2mm]\hline & 0_{5} \end{array}\right) & ,
\end{array}
\ee
the underlying gauge group being $\textrm{CSO}(5,0,3)$, dressed up with a further non-semisimple extension due to the presence of 7-form gauge flux, in analogy with the twisted $T^{7}$ case.

\item {\bf Model 2 \& 3:} The decomposition required in these cases is 
\be
\begin{array}{lclc}
\textrm{SL}(8)  & \supset  & \left(\mathbb{R}^{+}\right)^{5} \times \textrm{SL}(2)_{a} \times \textrm{SL}(2)_{i}  & , 
\end{array}
\ee
where three suitable and uniquely determined linear combinations of the five $\mathbb{R}^{+}$ charges introduced above represent the STU-charges. As already anticipated earlier, the resulting 
superpotentials are non-isotropic and their explicit form is beyond the present scope. Nevertheless, the qualitative analysis of the STU couplings induced by fluxes in these models are respectively 
collected in tables~\ref{Table:Fluxes_S4T3_2} and \ref{Table:Fluxes_S4T3_3}.
\begin{table}[t!]
\begin{center}
\begin{tabular}{|c|c|c|c|}
\hline
STU couplings & M-theory fluxes & Flux labels & $\mathbb{R}^{+}_{S} \times \mathbb{R}^{+}_{T} \times \mathbb{R}^{+}_{U} \times \textrm{SL}(2)_{a} \times \textrm{SL}(2)_{i}$ irrep's \\
\hline
\hline
$1$ & $G_{aibjck7}$ & $a_{0}$ & $(\textbf{1},\textbf{1})_{(+\frac{1}{2};+\frac{3}{2};+\frac{3}{2})}$ \\
\hline
\hline
$T$ & $G_{ibc7}$ & $c_{0}$ & $(\textbf{1},\textbf{1})_{(+\frac{1}{2};+\frac{1}{2};+\frac{3}{2})}$ \\
\hline
\hline
$T \, U$  & ${\omega_{a j}}^{k}$  & $c_{1}$  & $(\textbf{1},\textbf{3})_{(+\frac{1}{2};+\frac{1}{2};+\frac{1}{2})}$\\
\hline
$S \, U$ & ${\omega_{jk}}^{a}$ & $ b_{1}$ & $(\textbf{1},\textbf{1})_{(-\frac{1}{2};+\frac{3}{2};+\frac{1}{2})}$ \\
\hline
\hline
$S\,U^{2}$ & $\Theta_{ii}$ &  $b_{2}$ & $(\textbf{1},\textbf{1})_{(-\frac{1}{2};+\frac{3}{2};-\frac{1}{2})}$\\
\hline
$S\,T^{2}$ & $\Theta_{77}$ &  $d_{3}^{\prime}$ & $(\textbf{1},\textbf{1})_{(-\frac{1}{2};-\frac{1}{2};+\frac{3}{2})}$\\
\hline
$T\,U^{2} $ & $\Theta_{bc}$ & $ \tilde{c}_{1}^{\prime}$ & $(\textbf{3},\textbf{1})_{(+\frac{1}{2};+\frac{1}{2};-\frac{1}{2})}$ \\
\hline
$S\,T\,U \,$ & $\Theta_{i7}$ & $ \tilde{d}_{1}$ & $(\textbf{1},\textbf{1})_{(-\frac{1}{2};+\frac{1}{2};+\frac{1}{2})}$ \\
\hline
\end{tabular}
\end{center}
\caption{{\it Summary of M-theory fluxes and superpotential couplings on a twisted $S^{4}\times T^{3}$ according to Model 2. 
Please note that there appear several cubic couplings producing an intrinsically non-isotropic model.}}
\label{Table:Fluxes_S4T3_2}
\end{table}

\begin{table}[t!]
\begin{center}
\begin{tabular}{|c|c|c|c|}
\hline
STU couplings & M-theory fluxes & Flux labels & $\mathbb{R}^{+}_{S} \times \mathbb{R}^{+}_{T} \times \mathbb{R}^{+}_{U} \times \textrm{SL}(2)_{a} \times \textrm{SL}(2)_{i}$ irrep's \\
\hline
\hline
$1$ & $G_{aibjck7}$ & $a_{0}$ & $(\textbf{1},\textbf{1})_{(+\frac{1}{2};+\frac{3}{2};+\frac{3}{2})}$ \\
\hline
\hline
$U$ & $ G_{bjck}$ & $a_{1}$ & $(\textbf{1},\textbf{1})_{(+\frac{1}{2};+\frac{3}{2};+\frac{1}{2})}$\\
\hline
\hline
$S \, T$ & ${\omega_{7i}}^{a}$ & $d_{0}$ & $(\textbf{1},\textbf{1})_{(-\frac{1}{2};+\frac{1}{2};+\frac{3}{2})}$\\
\hline
$T^{2}$ & ${\omega_{a7}}^{i}$ &  $c_{3}^{\prime}$ & $(\textbf{1},\textbf{1})_{(+\frac{1}{2};-\frac{1}{2};+\frac{3}{2})}$\\
\hline
$U^{2} $ & ${\omega_{ai}}^{7}$ & $a_{2}$ & $(\textbf{1},\textbf{1})_{(+\frac{1}{2};+\frac{3}{2};-\frac{1}{2})}$ \\
\hline
\hline
$T^{2} \, U$ & $\Theta_{bc}$ & $ \tilde{c}_{2}^{\prime}$ & $(\textbf{3},\textbf{1})_{(+\frac{1}{2};-\frac{1}{2};+\frac{1}{2})}$ \\
\hline
$S \, T \, U $ & $\Theta_{jk}$ & $\tilde{d}_{1}$ & $(\textbf{1},\textbf{3})_{(-\frac{1}{2};+\frac{1}{2};+\frac{1}{2})}$ \\
\hline
\end{tabular}
\end{center}
\caption{{\it Summary of M-theory fluxes and superpotential couplings on a twisted $S^{4}\times T^{3}$ according to Model 3. 
Please note that there appeare several cubic couplings producing an intrinsically non-isotropic model.}}
\label{Table:Fluxes_S4T3_3}
\end{table}

\end{itemize}

\section{Discussion}

In this paper we have considered compactifications of M-theory on manifolds with non-trivial topologies. After reviewing the twisted $T^{7}$ and the $S^{7}$ cases, we also analysed the $S^{4} \times T^{3}$ case. By means of a group-theoretical approach, we have derived the flux-induced superpotentials in all the different cases in question. While the twisted $T^{7}$ superpotential contains terms that are at most quadratic in the complex scalars, the $S^{7}$ and $S^{4} \times T^{3}$, contain some quartic and cubic terms, respectively.

The appearence of the aforementioned higher-degree superpotential couplings may be na\"ively judged as a sign of non-geometry if one insists on a toroidal interpretation of the corresponding M-theory background. With these examples we show how non-geometric backgrounds still satisfying the section constraint can have a globally geometric eleven-dimensional origin from compactifications on topologies other than toroidal.
In this perspective it is not surprising that we could find all kinds of cubic couplings except the $U^{3}$, which would correspond to turning on the Romans' mass after reduction on a circle down to type IIA. This is in line with the statement that the mass parameter cannot be written as a derivative of any gauge field in a generalised geometry language \cite{Aldazabal:2010ef}. 

Nevertheless, whether or not such models actually contain new physics needs to be checked case by case. The origin of this open question is to be found in the local equivalence between different reformulations of eleven-dimensional supergravity that relate inequivalent solutions to the section condition. At a global level, the different and locally equivalent Lagrangians may differ by total derivative terms that might become important upon dimensional reduction. Such a fact has been already investigated in ten dimensions in the context of the so-called $\beta$-supergravity \cite{Andriot:2013xca,Andriot:2014qla}.

In particular, as far as (meta)stable dS extrema are concerned, all the examples known so far generically violate the section condition, thus being genuinely non-geometric. A possible future issue to be addressed is the existence of such (meta)stable dS vacua within locally geometric backgrounds that can be made globally geometric by following our approach.

Another possible line of research left open by our analysis is the possible relevance of exotic differentiable structures on spheres. The idea that our approach might capture some information about those comes from observing that the expression of the scalar potential associated with the $S^{7}$ reduction contains volume scaling behaviours that are beyond the ones predicted by its ordinary Riemannian structure.
In ref.~\cite{Englert:1982vs}, it was already observed that M-theory solutions on the $S^{7}$ seem to require other parallelisable differentiable structures beyond the Riemannian one. This might be
seen as an evidence that maximal supersymmetry and 11D supergravity are in fact sensitive to exotic differentiable structures on $S^{7}$. If this turns out to be the case, one could imagine using
M-theory reductions, and their underlying lower-dimensional gauged supergravity descriptions, to test the presence of exotic differentiable structures in other cases of special mathematical interest like, \emph{e.g.}, $S^{4}$.

\section*{Acknowledgments}

We would like to thank W.~H.~Baron, G.~Dall'Agata, T.~Ekholm, and M.~Larfors for very interesting and stimulating discussions.
The work of the authors was supported by the Swedish Research Council (VR). 

%
%

\appendix

\section{Relevant Branching Rules}
\label{sec:app1}

In this appendix we collect the whole set of branching rules used in the present paper. We refer to \cite{Feger:2012bs} for the conventions adopted
here.

\be
\begin{array}{cclc}
\textrm{E}_{7(7)}  & \supset  & \textrm{SL}(8)  & , \\[3mm]
\textbf{56}  & \rightarrow  & \textbf{28}  \oplus \textbf{28}^{\prime}  & ,\\[2mm]
\textbf{912}  & \rightarrow  & \textbf{36} \oplus \textbf{36}^{\prime}  \oplus \textbf{420} \oplus \textbf{420}^{\prime}  & .
\end{array}\label{E7->SL8}
\ee

\be
\begin{array}{cclc}
\textrm{SL}(8)  & \supset  & \mathbb{R}^{+}_{M} \times \textrm{SL}(7)  & , \\[3mm]
\textbf{28}  & \rightarrow  & \textbf{7}_{(-6)} \oplus \textbf{21}_{(+2)} & ,\\[2mm]
\textbf{36}  & \rightarrow  & \textbf{1}_{(-14)}  \oplus \textbf{7}_{(-6)} \oplus \textbf{28}_{(+2)} & ,\\[2mm]
\textbf{420}  & \rightarrow  & \textbf{21}_{(+2)}  \oplus \textbf{35}_{(+10)} \oplus \textbf{140}_{(-6)} \oplus \textbf{224}_{(+2)} & ,
\end{array}
\ee
where the subscripts in the above decompisotions denote $\mathbb{R}^{+}_{M}$ charges.

\be
\begin{array}{cclc}
\textrm{SL}(7)  & \supset  & \mathbb{R}^{+}_{B} \times \textrm{SL}(6)  & , \\[3mm]
\textbf{7}  & \rightarrow  & \textbf{1}_{(-6)}  \oplus \textbf{6}_{(+1)} & ,\\[2mm]
\textbf{21}  & \rightarrow  & \textbf{6}_{(-5)} \oplus \textbf{15}_{(+2)} & ,\\[2mm]
\textbf{28}  & \rightarrow  & \textbf{1}_{(-12)}  \oplus \textbf{6}_{(-5)} \oplus \textbf{21}_{(+2)} & ,\\[2mm]
\textbf{35}  & \rightarrow  & \textbf{15}_{(-4)}  \oplus \textbf{20}_{(+3)} & ,\\[2mm]
\textbf{140}  & \rightarrow  & \textbf{6}_{(+1)}  \oplus \textbf{15}_{(+8)} \oplus \textbf{35}_{(-6)} \oplus \textbf{84}_{(+1)} & ,\\[2mm]
\textbf{224}  & \rightarrow  & \textbf{15}_{(+2)}  \oplus \textbf{20}_{(+9)} \oplus \textbf{84}_{(-5)} \oplus \textbf{105}_{(+2)} & ,
\end{array}
\ee
where the subscripts in the above decompositions denote $\mathbb{R}^{+}_{B}$ charges.

\be
\begin{array}{cclc}
\textrm{SL}(6)  & \supset  & \mathbb{R}^{+}_{A} \times \textrm{SL}(3)_{a} \times \textrm{SL}(3)_{i}  & , \\[3mm]
\textbf{6}  & \rightarrow  & (\textbf{3},\textbf{1})_{(+1)}  \oplus (\textbf{1},\textbf{3})_{(-1)} & ,\\[2mm]
\textbf{15}  & \rightarrow  & (\textbf{3}^{\prime},\textbf{1})_{(+2)}  \oplus (\textbf{1},\textbf{3}^{\prime})_{(-2)} \oplus (\textbf{3},\textbf{3})_{(0)} & ,\\[2mm]
\textbf{20}  & \rightarrow  & (\textbf{1},\textbf{1})_{(+3)}  \oplus (\textbf{1},\textbf{1})_{(-3)} \oplus (\textbf{3},\textbf{3}^{\prime})_{(-1)}  \oplus (\textbf{3}^{\prime},\textbf{3})_{(+1)} & ,\\[2mm]
\textbf{21}  & \rightarrow  & (\textbf{3},\textbf{3})_{(0)}  \oplus (\textbf{6},\textbf{1})_{(+2)}  \oplus (\textbf{1},\textbf{6})_{(-2)} & ,\\[2mm]
\textbf{35}  & \rightarrow  & (\textbf{1},\textbf{1})_{(0)}  \oplus (\textbf{3},\textbf{3}^{\prime})_{(+2)}  \oplus (\textbf{3}^{\prime},\textbf{3})_{(-2)}  \oplus (\textbf{8},\textbf{1})_{(0)}  \oplus (\textbf{1},\textbf{8})_{(0)}  & ,\\[2mm]
\textbf{84}  & \rightarrow  & (\textbf{3},\textbf{1})_{(+1)}  \oplus (\textbf{1},\textbf{3})_{(-1)} \oplus (\textbf{3}^{\prime},\textbf{3}^{\prime})_{(+3)}  \oplus (\textbf{3}^{\prime},\textbf{3}^{\prime})_{(-3)} \oplus & \\
 &  &   (\textbf{6}^{\prime},\textbf{1})_{(+1)}  \oplus (\textbf{1},\textbf{6}^{\prime})_{(-1)} \oplus (\textbf{8},\textbf{3})_{(-1)}  \oplus (\textbf{3},\textbf{8})_{(+1)} & ,\\[2mm]
\textbf{105}  & \rightarrow  & (\textbf{3}^{\prime},\textbf{1})_{(+2)}  \oplus (\textbf{1},\textbf{3}^{\prime})_{(-2)}  \oplus (\textbf{3}^{\prime},\textbf{1})_{(-4)}  \oplus (\textbf{1},\textbf{3}^{\prime})_{(+4)} \oplus (\textbf{3},\textbf{3})_{(0)} \oplus & \\
 &  &   (\textbf{3},\textbf{6}^{\prime})_{(0)}  \oplus (\textbf{6}^{\prime},\textbf{3})_{(0)} \oplus (\textbf{3}^{\prime},\textbf{8})_{(+2)}  \oplus (\textbf{8},\textbf{3}^{\prime})_{(-2)} & ,
\end{array}
\ee
where the subscripts in the above decompositions denote $\mathbb{R}^{+}_{A}$ charges.

\be
\begin{array}{cclc}
\textrm{SL}(8)  & \supset  & \mathbb{R}^{+}_{Q} \times \textrm{SL}(3)_{a} \times \textrm{SL}(5)  & , \\[3mm]
\textbf{28}  & \rightarrow  & (\textbf{3}^{\prime},\textbf{1})_{(-10)}  \oplus (\textbf{3},\textbf{5})_{(-2)} \oplus (\textbf{1},\textbf{10})_{(+6)}  & ,\\[2mm]
\textbf{36}  & \rightarrow  & (\textbf{6},\textbf{1})_{(-10)}  \oplus (\textbf{3},\textbf{5})_{(-2)} \oplus (\textbf{1},\textbf{15})_{(+6)}  & ,\\[2mm]
\textbf{420}  & \rightarrow  & (\textbf{3}^{\prime},\textbf{1})_{(-10)} \oplus (\textbf{1},\textbf{5}^{\prime})_{(-18)} \oplus  (\textbf{3},\textbf{5})_{(-2)} \oplus (\textbf{1},\textbf{10})_{(+6)} \oplus (\textbf{6}^{\prime},\textbf{5})_{(-2)} \oplus & \\
  &   &  (\textbf{3}^{\prime},\textbf{10}^{\prime})_{(+14)} \oplus (\textbf{8},\textbf{10})_{(+6)} \oplus  (\textbf{3}^{\prime},\textbf{24})_{(-10)} \oplus (\textbf{1},\textbf{40})_{(+6)} \oplus (\textbf{3},\textbf{45})_{(-2)} & ,
\end{array}
\ee
where the subscripts in the above decompisotions denote $\mathbb{R}^{+}_{Q}$ charges.

\be
\begin{array}{cclc}
\textrm{SL}(5)  & \supset  & \mathbb{R}^{+}_{1} \times \textrm{SL}(4)  & , \\[3mm]
\textbf{5}  & \rightarrow  & \textbf{1}_{(-4)}  \oplus \textbf{4}_{(+1)}  & ,\\[2mm]
\textbf{10}  & \rightarrow  & \textbf{4}_{(-3)}  \oplus \textbf{6}_{(+2)} & ,\\[2mm]
\textbf{15}  & \rightarrow  & \textbf{1}_{(-8)}  \oplus \textbf{4}_{(-3)} \oplus \textbf{10}_{(+2)} & ,\\[2mm]
\textbf{24}  & \rightarrow  & \textbf{1}_{(0)}  \oplus \textbf{4}_{(+5)} \oplus \textbf{4}^{\prime}_{(-5)} \oplus \textbf{15}_{(0)} & ,\\[2mm]
\textbf{40}  & \rightarrow  & \textbf{4}^{\prime}_{(+7)}  \oplus \textbf{6}_{(+2)} \oplus \textbf{10}^{\prime}_{(+2)} \oplus \textbf{20}_{(-3)} & ,\\[2mm]
\textbf{45}  & \rightarrow  & \textbf{4}_{(+1)}  \oplus \textbf{6}_{(+6)} \oplus \textbf{15}_{(-4)} \oplus \textbf{20}_{(+1)} & ,
\end{array}
\ee
where the subscripts in the above decompisotions denote $\mathbb{R}^{+}_{1}$ charges.

\be
\begin{array}{cclc}
\textrm{SL}(4)  & \supset  & \mathbb{R}^{+}_{2} \times \textrm{SL}(3)_{i}  & , \\[3mm]
\textbf{4}  & \rightarrow  & \textbf{1}_{(-3)}  \oplus \textbf{3}_{(+1)}  & ,\\[2mm]
\textbf{6}  & \rightarrow  & \textbf{3}_{(-2)}  \oplus \textbf{3}^{\prime}_{(+2)} & ,\\[2mm]
\textbf{10}  & \rightarrow  & \textbf{1}_{(-6)}  \oplus \textbf{3}_{(-2)} \oplus \textbf{6}_{(+2)} & ,\\[2mm]
\textbf{15}  & \rightarrow  & \textbf{1}_{(0)}  \oplus \textbf{3}_{(+4)} \oplus \textbf{3}^{\prime}_{(-4)} \oplus \textbf{8}_{(0)} & ,\\[2mm]
\textbf{20}  & \rightarrow  & \textbf{3}_{(+1)}  \oplus \textbf{3}^{\prime}_{(+5)}  \oplus \textbf{6}^{\prime}_{(+1)} \oplus \textbf{8}_{(-3)} & ,
\end{array}
\ee
where the subscripts in the above decompisotions denote $\mathbb{R}^{+}_{2}$ charges.

%
%

\small

\clearpage

\bibliography{references}

\providecommand{\href}[2]{#2}\begingroup\raggedright\begin{thebibliography}{10}

\bibitem{Hertzberg:2007wc}
M.~P. Hertzberg, S.~Kachru, W.~Taylor, and M.~Tegmark, ``{Inflationary
  Constraints on Type IIA String Theory},''
  \href{http://dx.doi.org/10.1088/1126-6708/2007/12/095}{{\em JHEP} {\bf 0712}
  (2007)  095},
\href{http://arxiv.org/abs/0711.2512}{{\tt arXiv:0711.2512 [hep-th]}}.

\bibitem{Caviezel:2008tf}
C.~Caviezel, P.~Koerber, S.~Kors, D.~Lust, T.~Wrase, {\em et al.}, ``{On the
  Cosmology of Type IIA Compactifications on SU(3)-structure Manifolds},''
  \href{http://dx.doi.org/10.1088/1126-6708/2009/04/010}{{\em JHEP} {\bf 0904}
  (2009)  010},
\href{http://arxiv.org/abs/0812.3551}{{\tt arXiv:0812.3551 [hep-th]}}.

\bibitem{Danielsson:2011au}
U.~H. Danielsson, S.~S. Haque, P.~Koerber, G.~Shiu, T.~Van~Riet, {\em et al.},
  ``{De Sitter hunting in a classical landscape},''
  \href{http://dx.doi.org/10.1002/prop.201100047}{{\em Fortsch.Phys.} {\bf 59}
  (2011)  897--933},
\href{http://arxiv.org/abs/1103.4858}{{\tt arXiv:1103.4858 [hep-th]}}.

\bibitem{Dibitetto:2011gm}
G.~Dibitetto, A.~Guarino, and D.~Roest, ``{Charting the landscape of N=4 flux
  compactifications},'' \href{http://dx.doi.org/10.1007/JHEP03(2011)137}{{\em
  JHEP} {\bf 1103} (2011)  137},
\href{http://arxiv.org/abs/1102.0239}{{\tt arXiv:1102.0239 [hep-th]}}.

\bibitem{Dall'Agata:2005fm}
G.~Dall'Agata and N.~Prezas, ``{Scherk-Schwarz reduction of M-theory on
  G2-manifolds with fluxes},''
  \href{http://dx.doi.org/10.1088/1126-6708/2005/10/103}{{\em JHEP} {\bf 0510}
  (2005)  103},
\href{http://arxiv.org/abs/hep-th/0509052}{{\tt arXiv:hep-th/0509052
  [hep-th]}}.

\bibitem{Derendinger:2014wwa}
J.-P. Derendinger and A.~Guarino, ``{A second look at gauged supergravities
  from fluxes in M-theory},''
  \href{http://dx.doi.org/10.1007/JHEP09(2014)162}{{\em JHEP} {\bf 1409} (2014)
   162},
\href{http://arxiv.org/abs/1406.6930}{{\tt arXiv:1406.6930 [hep-th]}}.

\bibitem{Shelton:2005cf}
J.~Shelton, W.~Taylor, and B.~Wecht, ``{Nongeometric flux compactifications},''
  \href{http://dx.doi.org/10.1088/1126-6708/2005/10/085}{{\em JHEP} {\bf 0510}
  (2005)  085},
\href{http://arxiv.org/abs/hep-th/0508133}{{\tt arXiv:hep-th/0508133
  [hep-th]}}.

\bibitem{Baron:2014bya}
W.~H. Baron and G.~Dall'Agata, ``{Uplifting non-compact gauged
  supergravities},''
\href{http://arxiv.org/abs/1410.8823}{{\tt arXiv:1410.8823 [hep-th]}}.

\bibitem{Berman:2011cg}
D.~S. Berman, H.~Godazgar, M.~Godazgar, and M.~J. Perry, ``{The Local
  symmetries of M-theory and their formulation in generalised geometry},''
  \href{http://dx.doi.org/10.1007/JHEP01(2012)012}{{\em JHEP} {\bf 1201} (2012)
   012},
\href{http://arxiv.org/abs/1110.3930}{{\tt arXiv:1110.3930 [hep-th]}}.

\bibitem{Hohm:2013vpa}
O.~Hohm and H.~Samtleben, ``{Exceptional Field Theory I: $E_{6(6)}$ covariant
  Form of M-Theory and Type IIB},''
  \href{http://dx.doi.org/10.1103/PhysRevD.89.066016}{{\em Phys.Rev.} {\bf D89}
  (2014)  066016},
\href{http://arxiv.org/abs/1312.0614}{{\tt arXiv:1312.0614 [hep-th]}}.

\bibitem{Hohm:2013uia}
O.~Hohm and H.~Samtleben, ``{Exceptional Field Theory II: E$_{7(7)}$},''
  \href{http://dx.doi.org/10.1103/PhysRevD.89.066017}{{\em Phys.Rev.} {\bf D89}
  (2014)  066017},
\href{http://arxiv.org/abs/1312.4542}{{\tt arXiv:1312.4542 [hep-th]}}.

\bibitem{Hohm:2014qga}
O.~Hohm and H.~Samtleben, ``{Consistent Kaluza-Klein Truncations via
  Exceptional Field Theory},''
\href{http://arxiv.org/abs/1410.8145}{{\tt arXiv:1410.8145 [hep-th]}}.

\bibitem{Andriot:2012wx}
D.~Andriot, O.~Hohm, M.~Larfors, D.~Lust, and P.~Patalong, ``{A geometric
  action for non-geometric fluxes},''
  \href{http://dx.doi.org/10.1103/PhysRevLett.108.261602}{{\em Phys.Rev.Lett.}
  {\bf 108} (2012)  261602},
\href{http://arxiv.org/abs/1202.3060}{{\tt arXiv:1202.3060 [hep-th]}}.

\bibitem{Dibitetto:2012rk}
G.~Dibitetto, J.~Fernandez-Melgarejo, D.~Marques, and D.~Roest, ``{Duality
  orbits of non-geometric fluxes},''
  \href{http://dx.doi.org/10.1002/prop.201200078}{{\em Fortsch.Phys.} {\bf 60}
  (2012)  1123--1149},
\href{http://arxiv.org/abs/1203.6562}{{\tt arXiv:1203.6562 [hep-th]}}.

\bibitem{Musaev:2014lna}
E.~Musaev and H.~Samtleben, ``{Fermions and Supersymmetry in $\rm E_{6(6)}$
  Exceptional Field Theory},''
\href{http://arxiv.org/abs/1412.7286}{{\tt arXiv:1412.7286 [hep-th]}}.

\bibitem{deCarlos:2009qm}
B.~de~Carlos, A.~Guarino, and J.~M. Moreno, ``{Complete classification of
  Minkowski vacua in generalised flux models},''
  \href{http://dx.doi.org/10.1007/JHEP02(2010)076}{{\em JHEP} {\bf 1002} (2010)
   076},
\href{http://arxiv.org/abs/0911.2876}{{\tt arXiv:0911.2876 [hep-th]}}.

\bibitem{Danielsson:2012by}
U.~Danielsson and G.~Dibitetto, ``{On the distribution of stable de Sitter
  vacua},'' \href{http://dx.doi.org/10.1007/JHEP03(2013)018}{{\em JHEP} {\bf
  1303} (2013)  018},
\href{http://arxiv.org/abs/1212.4984}{{\tt arXiv:1212.4984 [hep-th]}}.

\bibitem{Blaback:2013ht}
J.~Bl{\aa}b{\"a}ck, U.~Danielsson, and G.~Dibitetto, ``{Fully stable dS vacua
  from generalised fluxes},''
  \href{http://dx.doi.org/10.1007/JHEP08(2013)054}{{\em JHEP} {\bf 1308} (2013)
   054},
\href{http://arxiv.org/abs/1301.7073}{{\tt arXiv:1301.7073 [hep-th]}}.

\bibitem{Maldacena:2000mw}
J.~M. Maldacena and C.~Nunez, ``{Supergravity description of field theories on
  curved manifolds and a no go theorem},''
  \href{http://dx.doi.org/10.1142/S0217751X01003937}{{\em Int.J.Mod.Phys.} {\bf
  A16} (2001)  822--855},
\href{http://arxiv.org/abs/hep-th/0007018}{{\tt arXiv:hep-th/0007018
  [hep-th]}}.

\bibitem{deWit:2007mt}
B.~de~Wit, H.~Samtleben, and M.~Trigiante, ``{The Maximal D=4
  supergravities},''
  \href{http://dx.doi.org/10.1088/1126-6708/2007/06/049}{{\em JHEP} {\bf 0706}
  (2007)  049},
\href{http://arxiv.org/abs/0705.2101}{{\tt arXiv:0705.2101 [hep-th]}}.

\bibitem{Dibitetto:2011eu}
G.~Dibitetto, A.~Guarino, and D.~Roest, ``{How to halve maximal
  supergravity},'' \href{http://dx.doi.org/10.1007/JHEP06(2011)030}{{\em JHEP}
  {\bf 1106} (2011)  030},
\href{http://arxiv.org/abs/1104.3587}{{\tt arXiv:1104.3587 [hep-th]}}.

\bibitem{Aldazabal:2008zza}
G.~Aldazabal, P.~G. Camara, and J.~Rosabal, ``{Flux algebra, Bianchi identities
  and Freed-Witten anomalies in F-theory compactifications},''
  \href{http://dx.doi.org/10.1016/j.nuclphysb.2009.01.006}{{\em Nucl.Phys.}
  {\bf B814} (2009)  21--52},
\href{http://arxiv.org/abs/0811.2900}{{\tt arXiv:0811.2900 [hep-th]}}.

\bibitem{Dall'Agata:2009gv}
G.~Dall'Agata, G.~Villadoro, and F.~Zwirner, ``{Type-IIA flux compactifications
  and N=4 gauged supergravities},''
  \href{http://dx.doi.org/10.1088/1126-6708/2009/08/018}{{\em JHEP} {\bf 0908}
  (2009)  018},
\href{http://arxiv.org/abs/0906.0370}{{\tt arXiv:0906.0370 [hep-th]}}.

\bibitem{Aldazabal:2006up}
G.~Aldazabal, P.~G. Camara, A.~Font, and L.~Ibanez, ``{More dual fluxes and
  moduli fixing},'' \href{http://dx.doi.org/10.1088/1126-6708/2006/05/070}{{\em
  JHEP} {\bf 0605} (2006)  070},
\href{http://arxiv.org/abs/hep-th/0602089}{{\tt arXiv:hep-th/0602089
  [hep-th]}}.

\bibitem{Dibitetto:2014sfa}
G.~Dibitetto, A.~Guarino, and D.~Roest, ``{Lobotomy of Flux
  Compactifications},'' \href{http://dx.doi.org/10.1007/JHEP05(2014)067}{{\em
  JHEP} {\bf 1405} (2014)  067},
\href{http://arxiv.org/abs/1402.4478}{{\tt arXiv:1402.4478 [hep-th]}}.

\bibitem{Danielsson:2012et}
U.~H. Danielsson, G.~Shiu, T.~Van~Riet, and T.~Wrase, ``{A note on obstinate
  tachyons in classical dS solutions},''
  \href{http://dx.doi.org/10.1007/JHEP03(2013)138}{{\em JHEP} {\bf 1303} (2013)
   138},
\href{http://arxiv.org/abs/1212.5178}{{\tt arXiv:1212.5178 [hep-th]}}.

\bibitem{Dibitetto:2012ia}
G.~Dibitetto, A.~Guarino, and D.~Roest, ``{Exceptional Flux
  Compactifications},'' \href{http://dx.doi.org/10.1007/JHEP05(2012)056}{{\em
  JHEP} {\bf 1205} (2012)  056},
\href{http://arxiv.org/abs/1202.0770}{{\tt arXiv:1202.0770 [hep-th]}}.

\bibitem{Freund:1980xh}
P.~G. Freund and M.~A. Rubin, ``{Dynamics of Dimensional Reduction},''
\href{http://dx.doi.org/10.1016/0370-2693(80)90590-0}{{\em Phys.Lett.} {\bf
  B97} (1980)  233--235}.

\bibitem{Englert:1982vs}
F.~Englert, ``{Spontaneous Compactification of Eleven-Dimensional
  Supergravity},''
\href{http://dx.doi.org/10.1016/0370-2693(82)90684-0}{{\em Phys.Lett.} {\bf
  B119} (1982)  339}.

\bibitem{DallAgata:2011aa}
G.~Dall'Agata and G.~Inverso, ``{On the Vacua of N = 8 Gauged Supergravity in 4
  Dimensions},'' \href{http://dx.doi.org/10.1016/j.nuclphysb.2012.01.023}{{\em
  Nucl.Phys.} {\bf B859} (2012)  70--95},
\href{http://arxiv.org/abs/1112.3345}{{\tt arXiv:1112.3345 [hep-th]}}.

\bibitem{Borghese:2012zs}
A.~Borghese, G.~Dibitetto, A.~Guarino, D.~Roest, and O.~Varela, ``{The
  SU(3)-invariant sector of new maximal supergravity},''
  \href{http://dx.doi.org/10.1007/JHEP03(2013)082}{{\em JHEP} {\bf 1303} (2013)
   082},
\href{http://arxiv.org/abs/1211.5335}{{\tt arXiv:1211.5335 [hep-th]}}.

\bibitem{Dibitetto:2012xd}
G.~Dibitetto, ``{Gauged Supergravities and the Physics of Extra Dimensions},''
\href{http://arxiv.org/abs/1210.2301}{{\tt arXiv:1210.2301 [hep-th]}}.

\bibitem{Samtleben:2005bp}
H.~Samtleben and M.~Weidner, ``{The Maximal D=7 supergravities},''
  \href{http://dx.doi.org/10.1016/j.nuclphysb.2005.07.028}{{\em Nucl.Phys.}
  {\bf B725} (2005)  383--419},
\href{http://arxiv.org/abs/hep-th/0506237}{{\tt arXiv:hep-th/0506237
  [hep-th]}}.

\bibitem{Blair:2014zba}
C.~D.~A. Blair and E.~Malek, ``{Geometry and fluxes of SL(5) exceptional field
  theory},''
\href{http://arxiv.org/abs/1412.0635}{{\tt arXiv:1412.0635 [hep-th]}}.

\bibitem{Aldazabal:2010ef}
G.~Aldazabal, E.~Andres, P.~G. Camara, and M.~Grana, ``{U-dual fluxes and
  Generalized Geometry},''
  \href{http://dx.doi.org/10.1007/JHEP11(2010)083}{{\em JHEP} {\bf 1011} (2010)
   083},
\href{http://arxiv.org/abs/1007.5509}{{\tt arXiv:1007.5509 [hep-th]}}.

\bibitem{Andriot:2013xca}
D.~Andriot and A.~Betz, ``{$\beta$-supergravity: a ten-dimensional theory with
  non-geometric fluxes, and its geometric framework},''
  \href{http://dx.doi.org/10.1007/JHEP12(2013)083}{{\em JHEP} {\bf 1312} (2013)
   083},
\href{http://arxiv.org/abs/1306.4381}{{\tt arXiv:1306.4381 [hep-th]}}.

\bibitem{Andriot:2014qla}
D.~Andriot and A.~Betz, ``{Supersymmetry with non-geometric fluxes, or a
  $\beta$-twist in Generalized Geometry and Dirac operator},''
\href{http://arxiv.org/abs/1411.6640}{{\tt arXiv:1411.6640 [hep-th]}}.

\bibitem{Feger:2012bs}
R.~Feger and T.~W. Kephart, ``{LieART - A Mathematica Application for Lie
  Algebras and Representation Theory},''
\href{http://arxiv.org/abs/1206.6379}{{\tt arXiv:1206.6379 [math-ph]}}.

\end{thebibliography}\endgroup
\bibliographystyle{utphys}

\end{document}